# A Perspective on Solar Energetic Particles

*Donald V. Reames* (https://orcid.org/0000-0001-9048-822X)

Institute for Physical Science and Technology, University of Maryland, College Park, MD, USA

dvreames@gmail.com

**Abstract** The author has been fortunate to observe and participate in the rise of the field of solar energetic particles (SEPs), from the early abundance studies, to the contemporary paradigm of shock acceleration in large SEP events, and element abundance enhancements that are power laws in mass-to-charge ratios from H to Pb. Through painful evolution the "birdcage" model and the "solar-flare myth" came and went, leaving us with shock waves and solar jets that can interact as sources of SEPs.

## Introduction

Often the evolution of science and of scientists seems a diffusive process, a random walk through topics and talented colleagues. It is common to think of planning a course of study, a proposal, or even an entire career in advance, but perhaps it works just as well with some randomness.

## Nuclear Emulsion

I became an undergraduate, then a graduate student, at the University of California at Berkeley. For my PhD thesis I studied nuclear interactions of heavy ions, especially O at 10 MeV amu$^{-1}$, using nuclear emulsion detectors under the guidance of emulsion expert Walter H. Barkas. At that time emulsions were also flown on balloons to study the composition of galactic cosmic rays (GCRs), and the first measurement of element abundances in solar energetic particles (SEPs) was made with emulsions on sounding rockets from Ft. Churchill, Manitoba by Fichtel and Guss (1961). I was impressed by these applications of emulsion to the budding Space Sciences and readily expressed my interest when Carl Fichtel contacted Barkas for possible new PhDs to hire. I began working at Goddard Space Flight Center in 1964, studying GCR abundances on the manned Gemini mission (Durgaprasad et al., 1970) and eventually extending the sounding rocket measurements of SEPs up to the element Fe (Bertsch et al., 1969).

During the 1970's, increasingly accurate SEP measurements began to be made almost continuously with *dE/dx* vs. *E* particle telescopes (measuring energy loss in a thin detector vs. total energy as a particle stops) on board satellites, eventually using Si solid-state detectors, and lower-resolution, labor-intensive nuclear emulsions faded from use. I marvel at how important they once were to my career: I became an astrophysicist because of my specialized knowledge of an obsolete technology which no longer exists. But by the time nuclear emulsions died I had become an astrophysicist. There was no turning back. By 1977 I was working on particle telescopes for spacecraft with Tycho von Rosenvinge and Frank McDonald.

## $^3$He-Rich Events

It was well known that GCRs fragment against interstellar H during their lifetime in space, $^4$He produces $^3$He and $^2$H while C, N, and O fragment to produce Li, Be, and B. Discovery of enhanced $^3$He in SEP abundances was first considered as new evidence for fragmentation in solar flares but soon came the discovery of ~1000-fold enhancements, $^3$He/$^4$He = 1.5 (Serlemitsos and Balasubrahmanyan, 1975) vs. 4 x 10$^{-4}$ in the solar wind, with no measurable $^2$H, Li, Be, or B at all. These events had nothing to do with fragmentation; this was a completely new resonance phenomenon, where waves resonate with the $^3$He gyrofrequency (e.g. Fisk, 1978; Temerin and Roth, 1992). These $^3$He-rich events were small (Mason, 2007). In fact, working with Robert Lin, we found that they were associated with small 10-100 keV electron events (Reames et al., 1985) and with type III radio bursts (Reames and Stone, 1986) that the streaming electrons produce. Years earlier, Wild et al. (1963) had distinguished two sources of SEPs producing different radio bursts (i) rapidly streaming electrons producing type III radio bursts and (ii) type II bursts producing shock waves known to accelerate ions as well. Lin (1970, 1974) had found that type III bursts often came from "pure" electron events. Apparently, these frequent type-III events were mostly $^3$He-rich events.

Many early theories of $^3$He enhancement (e.g. Fisk, 1978; see early references in Reames, 2021a, 2021c) involved selective heating of the $^3$He by resonant wave absorption followed by acceleration later by some unspecified mechanism. Attending a dinner following a spacecraft meeting, I was sitting across from Mike Temerin who asked "What do you do?" so I began to talk about these weird $^3$He-rich events. He thought there might be a relationship with the "ion conics" seen in the Earth's magnetosphere where mirroring ions absorb energy from impinging waves until they can finally escape, forming the conic spatial distribution. A result of this discussion was the paper by Temerin and Roth (1992) where streaming electrons produce the waves that are resonantly absorbed to preferentially accelerate $^3$He, providing both acceleration and a connection between the $^3$He and the abundant streaming type-III electrons. Ideas originate in many places.

These "impulsive" SEP events have associated enhancements of heavy elements which were eventually found to increase as the 3.6 power, on average, of an element's mass-to-charge ratio *A/Q*, its atomic mass *A* divided by its electronic charge *Q*, first up to Fe, and then throughout the periodic table by a factor of ~1000 between H and Pb (Reames and Ng, 2004; Mason et al., 2004; Reames et al., 2014a). Theoretical simulations associate these strong abundance enhancements with magnetic reconnection (Drake et al., 2009) on open field lines and an important association now connects impulsive SEP events with solar jets (Kahler et al., 2001; Reames, 2002; Nitta et al., 2006, 2015; Wang et al., 2006; Bučík et al., 2018a, 2018b ;Bučík, 2020). Escape on the open magnetic field lines from jets produces no nuclear fragments.

**CMEs**

Meanwhile, the large "gradual" SEP events were found by Kahler et al. (1984) to have a 96% correlation with fast, wide coronal mass ejections (CMEs) and the shock waves that they drive (Kahler, 2001; Gopalswamy et al., 2012; Kouloumvakos et al., 2019). Shock theory had been well developed for GCRs and was already being applied to "energetic storm particles" (ESPs) that peak at the interplanetary continuation of these same shocks (Lee, 1983, 2005). CME-driven shock acceleration explained the broad spatial distribution of large SEP events (Cane et al., 1988; Reames, 1999, 2013) replacing the "birdcage" model that was invented to allow protons to hop from loop to loop across the face of the corona from flares (see Reames 2021a). In contrast, shocks easily cross magnetic fields, accelerating particles over a broad front. The highest energies were produced by the shocks at the corona (Zank et al., 2000; Cliver et al., 2004; Ng and Reames 2008; Desai and Giacalone, 2016) near their onsets at ~2 $R_S$ (Reames, 2009a, 2009b). The work of Zank et al. (2000) led to the development of the iPATH models of SEP transport (e.g. Hu et al., 2018).

The growing realization of the importance of CMEs and of shock acceleration of SEPs, especially in large gradual events, was pointed out by Jack Gosling's (1993, 1994) paper "The solar-flare myth." This paper caused a great controversy with flare enthusiasts who had not followed the evolution of CME and SEP research (see Reames, 2021a or 2021b for relevant publications). We have now come to understand that flares do not contribute SEPs in space; flares are hot and bright precisely because all the energy from magnetic reconnection, including accelerated particles, is contained on closed magnetic loops or dumped into their footpoints, so only photons and neutrals escape (Mandzhavidze et al., 1999; Murphy et al., 1991, 2016). Jets are the open-field equivalents that act as a source of interplanetary SEPs (e.g. Bučík, 2020), but CME-driven shocks dominate large events.

As protons stream away from a shock they amplify Alfvén waves that scatter all ions coming behind. This strengthens the acceleration and scatters and traps lower-rigidity ions, limiting intensities at the "streaming limit" (Reames and Ng, 1998, 2010), flattening energy spectra (Reames and Ng, 2010), and altering element abundances (Reames et al., 2000). Study of this self-consistent theory of wave-particle interactions was led by Chee Keong Ng (Ng and Reames, 1994, 1995; Ng et al., 1999, 2003, 2012) and applied to the time evolution of shock acceleration (Ng and Reames, 2008). Hopefully, someone will continue and extend these careful self-consistent studies.

**FIP and *A/Q***

It had been known for many years (e.g. Webber, 1975) that the abundances of elements in SEPs had ~3x enhancements, relative to photospheric abundances, of elements with low (<10 eV) first ionization potential (FIP). This 3x enhancement is an ion-neutral fractionation during formation of the solar corona. Electromagnetic waves can affect low-FIP elements (e.g. Mg, Si, and Fe) that are initially ionized, but not high-FIP neutral atoms (e.g. O, Ne, and Ar) rising up to the corona where all become ionized. Incidentally, the FIP pattern of SEPs differs from that of the solar wind (Mewaldt et al., 2002; Reames 2018a; Laming et al., 2019); SEPs are not just accelerated solar wind. However, FIP may help locate the different sources of SEPs and solar wind in the corona (Brooks and Yardley 2021).

Meyer (1985; Reames, 1995, 2014) found that element abundances in SEP events, compared with photospheric abundances, consisted of a FIP effect, shared by all events, and a dependence upon *A/Q*, that varied from event to event. The FIP effect occurred during formation of the corona, while the *A/Q* dependence resulted during acceleration, much later. Breneman and Stone (1985) established a power-law dependence using average *Q* values measured by Luhn et al. (1984). However, the *Q* values of the ions depend upon source electron temperature as noted by Meyer (1985).

**Source Temperatures**

The relevance of temperature was also noted by Jean-Paul Meyer in impulsive SEP events (Reames et al., 1994) where $^4$He, C, N, and O abundances appeared un-enhanced because they were all fully ionized, while Ne, Mg, and Si had comparable enhancements because they were in stable two-electron states, while Fe was further enhanced. This configuration can occur at about 3 MK. Direct charge measurements in impulsive events had shown that elements up to Si were fully ionized (Luhn, 1987), thus they must then be stripped *after* acceleration, as was later proven (DiFabio et al., 2008).

Much later, we have been able to determine a temperatures for each event from its abundance measurements by trying $Q$ values for many temperatures to see which gives the best-fit power law of enhancements vs. $A/Q$ (Reames et al., 2014b); most impulsive SEP events yield ~2.5 MK and, recently, EUV temperatures in solar jet sources of impulsive events were also found to peak at ~2.5 MK (Bučík et al., 2021). Source temperatures for impulsive SEP events were mostly within the ~10% error of the determination, however, similar techniques applied to abundances of gradual SEP events (Reames, 2016a, 2018b) varied widely from 0.6 – 2 MK when dominated by ambient coronal ions and > 2 MK when they involved reaccelerated impulsive ions. These higher-temperature gradual SEP events fit in with the growing evidence that CME-driven shock waves could sometimes preferentially reaccelerate ions from residual impulsive suprathermal ions originally from jets (Desai et al., 2003; Tylka et al., 2001, 2005; Tylka and Lee, 2006; Sandroos and Vainio, 2007; Reames, 2016b). These suprathermal ions were found to collect in pools, perhaps from multiple small jets that are difficult to resolve (Desai et al., 2003; Wiedenbeck et al., 2008, 2013; Bučík et al., 2014, 2015; Chen et al., 2015) repeatedly sampled by shocks (Reames, 2022).

Clearly, SEPs now seemed more complicated than just impulsive events from jets and gradual events from CME-driven shocks. Kahler et al. (2001) found CMEs from the jets in impulsive events that could drive fast local shocks and there were also large CMEs in gradual events could sample pools of residual impulsive ions. Reames (2020) suggested four SEP classes: (1) SEP1 impulsive events from pure magnetic reconnection in jets, (2) SEP2 events with additional acceleration when the local CME from that jet is fast enough to produce a shock, (3) SEP3 events are dominated by seed particles from preexisting impulsive suprathermal pools that are traversed by wide, fast shocks, and (4) SEP4 events are accelerated by wide, fast shocks predominantly from the ambient corona. The new emphasis on shocks and jets was a major change from the previous "flare myth." The abundances of SEPs from impulsive events retain their unique signature even when combined with ambient plasma and reaccelerated by shock waves.

We knew about power-law dependence upon $A/Q$ in 1985 (Breneman and Stone, 1985). Powers in magnetic rigidity produce these powers in $A/Q$ at a given MeV amu$^{-1}$. We knew about the importance of $Q$ variations and temperature in determining abundances (Meyer, 1985; Luhn et al., 1985, 1987; Mason et al., 1985; Leske et al., 1995; Tylka et al., 1985; Reames et al., 1994). Yet it took ~20 years to relate this $A/Q$ dependence to source temperatures in impulsive (Reames et al., 2014b) and gradual (Reames, 2016a, 2018b) events and to shocks plying various seed populations. It is true that reaccelerated impulsive ions may have changed their $A/Q$ from stripping, but the patterns are dominated by their initial huge enhancement of the seed population while the $A/Q$ dependence in shock acceleration is weak.

**Perspectives**

Where did I learn astrophysics? Not in graduate school. Early in my career I acknowledge learning astrophysics theory from colleague Reuven Ramaty. I learned specifics about electrons from co-author Robert Lin, radio emission from Robert Stone, CMEs from Stephen Kahler, and detectors from Tycho von Rosenvinge. I learned by working with these and other colleagues. Later, I learned a great deal about particle acceleration and transport from many years of discussions with Chee Ng, but I also profited greatly by working with other co-authors, by reading papers, and by endlessly looking at data. I am still learning astrophysics.

The most important contributions to SEP studies, in my opinion, were the determinations that the source of gradual events is CME-driven shock waves (Kahler et al., 1984) and that the source of impulsive SEP events is reconnection in jets (Kahler et al., 2001; Bučík, 2020). A lot of early insights were overlooked: Wild et al. (1963) already knew about the two sources of SEPs; Meyer (1985, Figure 11) knew that source temperatures were an important determinant of abundances. Were flares taken so seriously just because they are easier to see than CMEs, shocks, or jets?

What is my most productive work? Ironically, an early review article (Reames, 1999) was not only well received as a first review of SEPs, but, it was especially helpful to me in collecting ideas that improved my own perspective. Thus, writing review articles can be as educational for the author as for the reader and I have written more as new areas evolved (Reames, 2013, 2015, 2018b, 2020, 2021b, 2021c). Textbooks are even better (Reames, 2021a). Regarding research articles, I think the recent articles on SEP temperatures cited above and the correlations of energy spectra with abundances (e.g. Reames, 2021d, 2022) will be as productive as the earlier articles on $^3$He-rich events and type III bursts, FIP, or onset times.

In recent years I have continued to work mostly with data from the LEMT on the *Wind* spacecraft, now 27 years of data. There are detailed spectra and element abundance measurements during hundreds of SEP events, all different, from this and many other spacecraft, all freely available on the web (https://cdaweb.gsfc.nasa.gov/sp_phys/). Yet there are so few other people who look at it that I seem to have exclusive access. In contrast, there are also armies of co-authors who flock to join a few select articles. Are these topics vastly more interesting? Am I missing something wonderful, or is the issue more about funding than scientific interest? Aye, there's the rub. We few retirees, funded only by pensions, are able to graze unmolested the choicest historic pastures of data from instruments that are yet unequalled – without even writing a proposal. When possible, find time to follow the physics, rather than the crowd.

Actually, proposals are also interesting. Can you predict what you will discover in the next three years? I cannot. Will you doggedly follow an approved plan even if a surprising new avenue suddenly opens? Some will. Many ideas sound good on paper but later turn out to be unsupported by the data. I once calculated my "batting average" as only slightly over .300. Should we publish all those losers, i.e. "good ideas" that did not work? Approved proposals can also suffer from "group think." It is not a perfect system but it is hard to suggest improvement – unless you are retired.

In my opinion, abundances are a key to underlying physics of SEPs that has been poorly exploited theoretically. Why are energy spectra correlated with abundance enhancements in "pure" (SEP4) shock events (e.g. Reames 2021d, 2022)? How can they then be completely independent in "pure" (SEP1) impulsive events? Where do resonances (e.g. $^3$He) fit into reconnection models that predict power laws in *A/Q*? Surely, there are opportunities for mirroring $^3$He in reconnection regions. Is C/O somehow suppressed, on average, in SEPs, or is it overestimated in the photosphere (e.g. Reames 2021b)? Are $^4$He/O depletions related to the high FIP of He; are there occasional He-poor regions in the solar corona (Reames 2017, 2019)? I am still trying to learn astrophysics.

We cannot produce beautiful images of the Sun with SEPs, but we have made significant progress with the data we do have. We have been doing "multi-messenger" science for 60 years with SEPs, type-II and type-III radio bursts, and CMEs, long before it became so fashionable. There is much more of it to do.

**Author Contributions**

All work on this article was performed by DVR

**Funding**

No institutional funding was provided for this work.

**Acknowledgments**

I thank my many colleagues who have helped advance the study of SEPs.